\documentclass{webofc}
\usepackage[varg]{txfonts}   
\usepackage{listings}
\usepackage{hyperref}
\newcommand{\fo}{\ifmmode \nu_{\rm 1O}\else$\nu_{\rm 1O}$\fi}
\newcommand{\fx}{\ifmmode \nu_{\rm x}\else$\nu_{\rm x}$\fi}
\newcommand{\Po}{\ifmmode P_{\rm 1O}\else$P_{\rm 1O}$\fi}
\newcommand{\Pf}{\ifmmode P_{\rm F}\else$P_{\rm F}$\fi}
\newcommand{\Px}{\ifmmode P_{\rm x}\else$P_{\rm x}$\fi}

\begin{document}
\title{Petersen Diagram Revolution}
%
%

\author{\firstname{Radoslaw} \lastname{Smolec}\inst{1}\fnsep\thanks{\href{mailto:smolec@camk.edu.pl}{\tt smolec@camk.edu.pl}} \and
        \firstname{Wojciech} \lastname{Dziembowski}\inst{1,2} \and 
        \firstname{Pawel} \lastname{Moskalik}\inst{1} \and 
        \firstname{Henryka} \lastname{Netzel}\inst{2,1} \and 
        \firstname{Zdenek} \lastname{Prudil}\inst{3} \and 
        \firstname{Marek} \lastname{Skarka}\inst{4} \and 
        \firstname{Igor} \lastname{Soszynski}\inst{2}
}

\institute{Nicolaus Copernicus Astronomical Center, Polish Academy of Sciences, ul. Bartycka 18, 00-716 Warszawa, Poland 
\and
           Warsaw University Observatory, Al. Ujazdowskie 4, 00-478 Warszawa, Poland
\and
           Astronomisches Rechen-Institut, Zentrum f\"ur Astronomie der Universit\"at Heidelberg, M\"onchhofstr. 12-14,
69120 Heidelberg, Germany
\and      Konkoly Observatory, Research Centre for Astronomy and Earth Sciences, Hungarian Academy of Sciences,
Konkoly Thege Mikl\'os \'ut 15-17, H-1121 Budapest, Hungary
          }

\abstract{%
 Over the recent years, the Petersen diagram for classical pulsators, Cepheids and RR~Lyr stars, populated with a few hundreds of new multiperiodic variables. We review our analyses of the OGLE data, which resulted in the significant extension of the known, and in the discovery of a few new and distinct forms of multiperiodic pulsation. The showcase includes not only radial mode pulsators, but also radial-non-radial pulsators and stars with significant modulation observed on top of the beat pulsation. First theoretical models explaining the new forms of stellar variability are briefly discussed. 
}
\maketitle
\section{Introduction}\label{sec:intro}

Classical pulsators, Cepheids and RR~Lyrae stars, are among the most important variables. As excellent standard candles, they are invaluable for distance determination and for studying the structure and evolution of nearby stellar systems. For years they were regarded as simple, typically single-periodic, radially pulsating stars. The only serious, unsolved problem was the Blazhko effect, quasi-periodic modulation of pulsation amplitude and/or phase, observed in a significant fraction of RR~Lyr stars. In the recent years, thanks to precise observations from space and thanks to huge amount of top-quality data gathered by the ground-based photometric sky surveys, many new dynamical phenomena were discovered in classical pulsators. In these proceedings, we focus on the new forms of multi-periodic pulsation. Most of the discoveries were done in the data collected by the Optical Gravitational Lensing Experiment (OGLE) \cite{OGLE}.

\section{Classical Cepheids}\label{sec:cep}

Several new classes of multi-mode radial pulsation in classical Cepheids were reported in the OGLE data, see \cite{o4_cep_multi} and references therein. These include triple-mode radial pulsators and rare double-mode radial pulsators with higher order overtones excited. In the most interesting group, in which the dominant variability is due to radial first overtone, and additional variability falls in the $\Px=(0.60,0.65)\Po$ range, non-radial pulsation must be involved. More than 200 such stars were reported e.g. in \cite{MK09, o3_cep_smc, o4_cep_multi}. The first detailed analysis of the largest sample of 138 SMC pulsators of the new type was presented in \cite{SS16}. In the Petersen diagram in Fig.~\ref{fig:pet} these stars are marked with filled squares and form three tight and close sequences, which immediately rules out the possibility that additional periodicities are due to consecutive radial overtones. Additional periodicities are always of low amplitude, in the mmag range. In the frequency spectrum, these periodicities manifest as a relatively broad power excess rather than a single and coherent peak. Most interestingly, in many stars, significant signal centered at a sub-harmonic frequency, $1/2\fx$, was detected. These signals were detected in 74 per cent of stars forming the middle sequence in the Petersen diagram and in 31 per cent of stars forming the top sequence. The signal was not detected in any star from the bottom sequence. This observation is crucial for the model explaining the nature of the additional variability, which we describe below, after discussing an analogous form of pulsation in RR~Lyr stars.

\section{RR~Lyr stars with the dominant radial first overtone}\label{sec:rrc}

\begin{figure*}
\centering
\includegraphics[width=7.cm,clip]{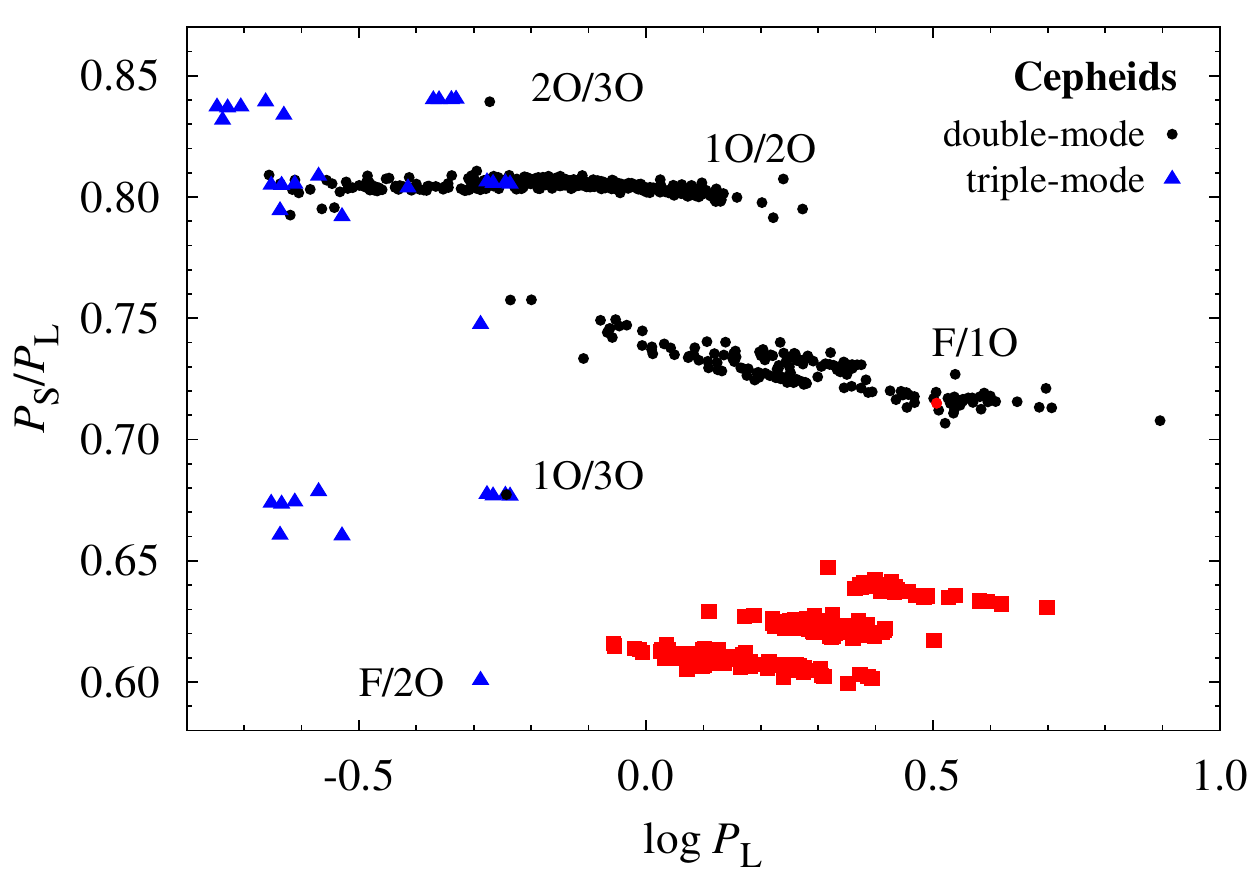}
\includegraphics[width=7.cm,clip]{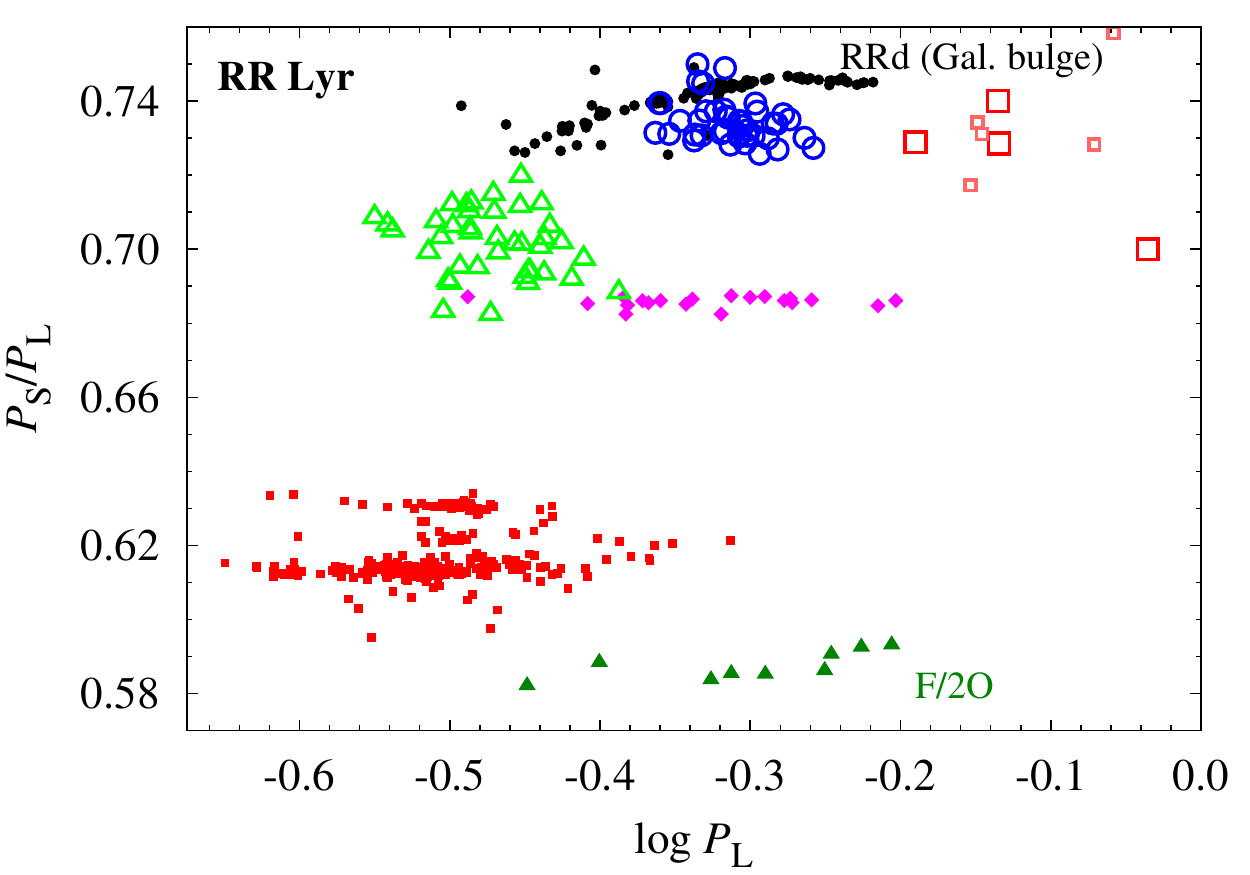}
\caption{Petersen diagram (shorter-to-longer period ratio vs. log of longer period) for classical Cepheids (left panel) and RR~Lyr stars (right panel) with various classes of multi-mode pulsators plotted with different symbols.}
\label{fig:pet}
\end{figure*}

{\bf RR$_{\boldsymbol{\rm 0.61}}$ stars}. AQ~Leo was the first RR~Lyr star in which additional variability with $\Px/\Po$ close to $0.61$ was detected \cite{aqleo}. Now we know more than three hundred RR~Lyr stars in which the first overtone is excited (either RRc or RRd) and in which additional variability in the $(0.60,0.64)\Po$ range was detected. The majority of these stars were discovered in the OGLE Galactic bulge collection \cite{NSM15a,NSM15b}. This form of pulsation must be common, as 14 out of 15 RRc/RRd stars observed at utmost precision from space show the additional variability, see \cite{NSM15b} and references therein. We denote these stars as RR$_{0.61}$. In the Petersen diagram in Fig.~\ref{fig:pet} they are marked with small filled squares. Three sequences are apparent, but not as distinct as in the case of Cepheids. The most populated sequence is centered at $\Px/\Po\approx 0.613$. Amplitude of the additional signal is low, typically 2 per cent of the first overtone amplitude. Just as in the case of Cepheids, in the frequency spectra of these stars, significant power at sub-harmonic frequency, $1/2\fx$, is often detected. Again, stars with this signal are not distributed randomly in the Petersen diagram: nearly all of them form the top sequence and only a few are among the stars forming the most numerous bottom sequence. Signal at $1/2\fx$ was not detected in stars forming the middle sequence. This least populated sequence arises because of stars in which three additional signals are simultaneously observed in the frequency spectrum. Interestingly, the middle signal is centered at the arithmetic average of the frequencies of the two remaining signals.

Recently, Dziembowski proposed a model explaining the nature of additional periodicity observed at $(0.60,\,0.65)\Po$ in Cepheids and in RR~Lyr stars \cite{WD16}. In this model, additional variability is due to non-radial pulsation modes with frequencies equal to $1/2\fx$ -- it is the signal at sub-harmonic frequency (which is not always detected) that corresponds to non-radial mode. These are acoustic, strongly trapped unstable modes of moderate degrees. Their degrees are $\ell\!=\!7$, $8$ and $9$ in classical Cepheids (top, middle and bottom sequence, respectively) and $\ell\!=\!8$ and $9$ in RR~Lyr stars (top and bottom sequence, respectively). The middle sequence in the case of RR~Lyr variables is due to combination frequency, $\nu_8+\nu_9$. Observed amplitude of a non-radial mode is reduced by geometric cancellation, which increases with increasing $\ell$, and, for moderate and high degrees, is stronger for odd $\ell$. Hence, there are higher chances to detect non-radial modes of $\ell\!=\!8$ than of $\ell\!=\!7$, and then, of $\ell\!=\!9$. This explains why signal at the sub-harmonic frequency is predominantly detected for the top sequence in the case of RR~Lyr stars and for the middle sequence in the case of Cepheids, as these sequences correspond to $\ell\!=\!8$. On the other hand, geometric and non-linear effects lead to high amplitude of the harmonic of the non-radial mode -- the signal observed at $\Px=(0.60,\,0.65)\Po$. 


{\bf Stars first reported by Netzel, Smolec \& Dziembowski, \cite{NSD15}}. This group was discovered in the OGLE data for Galactic bulge RR~Lyr stars. In eleven first overtone pulsators additional low-amplitude periodicity of longer period was discovered. Literature search revealed one additional star of the same type, KIC9453114, observed by {\it Kepler} and reported in \cite{PAMSM15}. Additional 8 stars were reported in \cite{NS16}. In Fig.~\ref{fig:pet} these stars are marked with filled diamonds. Period ratios tightly cluster at $\Po/\Px\!\approx\!0.686$, while first overtone periods cover a wide range from $\approx\!0.22$ to $\approx\!0.43$d. In all stars additional variability is coherent and of low amplitude. As this group is located well below the RRd sequence, period of the additional variability must be longer than the expected period of the fundamental mode in these stars. Hence, not only the additional variability cannot correspond to radial mode, but its interpretation in terms of non-radial modes is also difficult. As frequency is below the radial fundamental mode frequency, the supposed non-radial mode must be of gravity or of mixed character. Gravity modes are not expected to be excited in giant type stars. The nature of additional variability in these stars remains unexplained at the moment.

\section{RR~Lyr stars with the dominant radial fundamental mode}\label{sec:rrab}

{\bf Anomalous RRd stars.} Till recently, Blazhko effect was reported exclusively in single-periodic RR~Lyr stars. First detection of Blazhko effect in RRd pulsators was reported in the OGLE Galactic bulge data \cite{o4_rrl,SSU15} and in M3 observations \cite{JurcsikM3}. Later, Soszy\'nski et al. \cite{SSD16} reported similar stars in the OGLE Magellanic Cloud observations and defined a new class of {\it anomalous} RRd stars, which encompass the Blazhko RRd stars reported before. In the Petersen diagram in Fig.~\ref{fig:pet}, these stars are plotted with blue open circles. Their anomalous nature is revealed when the stars are compared to {\it classical} RRd stars, the double-mode pulsators plotted with black solid dots in Fig.~\ref{fig:pet}. The latter stars form a well defined and relatively tight progression in the Petersen diagram. Its short period, low period ratio tail is formed by metal-rich Galactic bulge RRd stars. There are three anomalous properties of the discussed stars. First, the period ratios are either too low, or, in a few cases, too high as compared to classical RRd stars of the same fundamental mode period. Still, these period ratios are easily explained as due to simultaneous pulsation in the radial fundamental and in radial first overtone modes. Second, unlike in classical RRd stars, in the majority of anomalous RRd stars the amplitude of the fundamental mode is higher than the amplitude of the first overtone. Third, in the majority of anomalous RRd stars, the periodic modulation of the pulsation modes is observed.

The cause of the anomalous period ratios and of the modulation, remains to be understood. In \cite{SSD16} we suggested that the parametric resonance between the three radial modes, $2\nu_{\rm 1O}\!=\!\nu_{\rm F}\!+\!\nu_{\rm 2O}$, might be essential in driving the pulsation of anomalous RRd stars, but dedicated study of this mechanism is needed. 

\begin{figure*}
\centering
\includegraphics[width=14cm,clip]{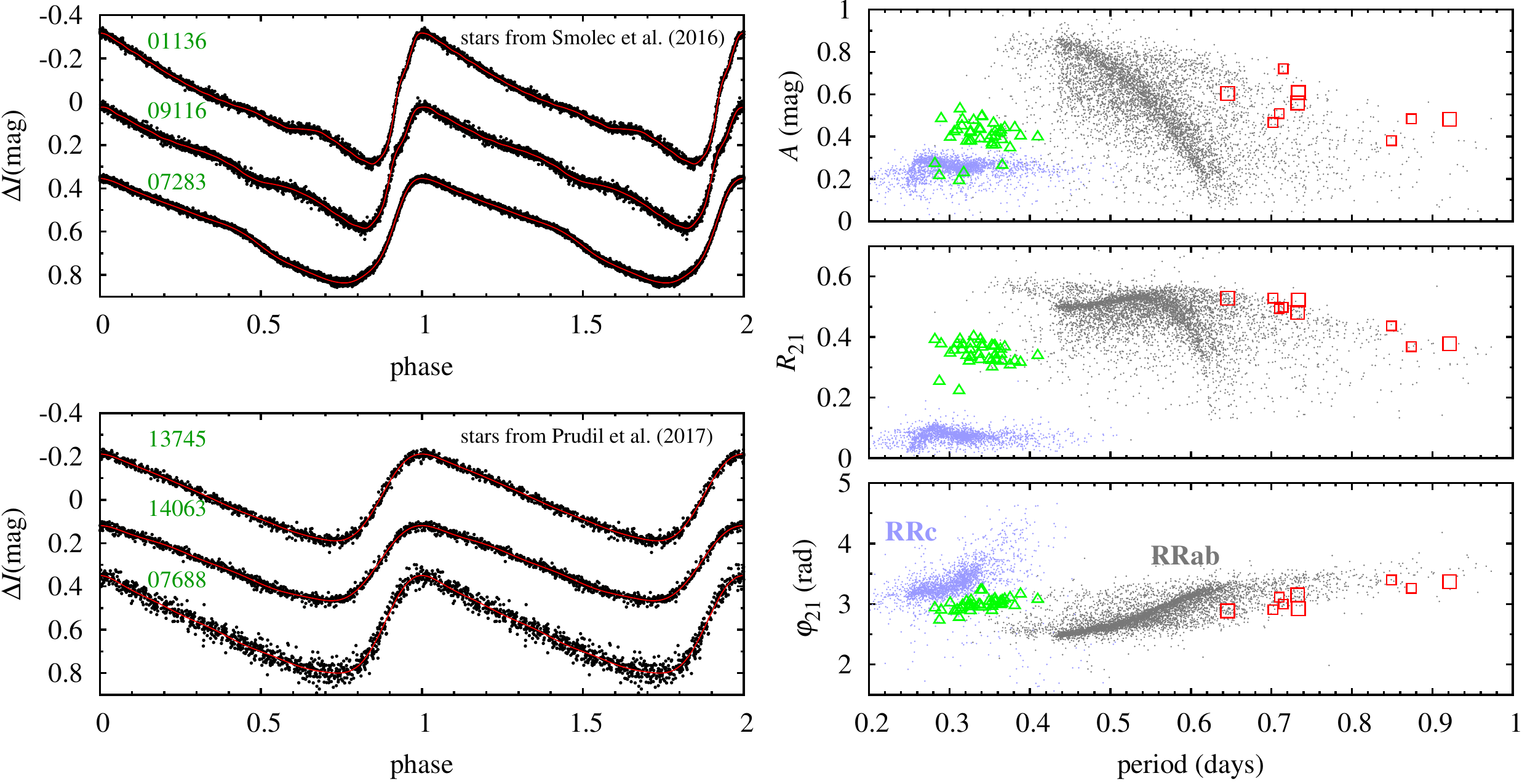}
\caption{Left panels: exemplary light curves corresponding to the dominant periodicity in stars from Smolec et al. \cite{SPSB16} (top) and Prudil et al. \cite{PSSN17} (bottom). Star's ids (OGLE-BLG-RRLYR-xxxxx) are given along each light curve. Right panel: Peak-to-peak amplitudes and low-order Fourier parameters ($R_{21}$ and $\varphi_{21}$) for all stars of these two groups. Symbols are the same as in Fig.~\ref{fig:pet}.}
\label{fig:lc}
\end{figure*}

{\bf Stars reported by Smolec et al. \cite{SPSB16}}. In a few long-period fundamental mode RR~Lyr stars of the OGLE Galactic bulge collection, additional variability of low amplitude and with a period shorter than the period of the radial fundamental mode was found. In the Petersen diagram, these stars are marked with open squares. The stars do not form any tight sequence in the Petersen diagram. The light curves corresponding to the dominant variability are among the typical for long-period fundamental mode RRab stars (Fig.~\ref{fig:lc}, top left), which is confirmed by the analysis of their Fourier parameters. Interestingly, the amplitudes and amplitude ratios are among the highest, while Fourier phases are among the lowest observed in RRab stars of similar period (Fig.~\ref{fig:lc}, right panels).

Linear pulsation calculations indicate that stars in this group might be extreme RRd stars, with the longest fundamental mode periods known. The only exception is OGLE-BLG-RRLYR-07283. Its low period ratio, nearly exactly equal to $0.7$ ($\Px/\Pf=0.700008$), cannot be fit with the models. This star has the largest amplitude of the additional variability, $8.5$ per cent of the radial mode amplitude.


{\bf Stars reported by Prudil et al. \cite{PSSN17}}. Another group of double-periodic pulsators is located at the short period extension of the classical RRd sequence. In 42 RR~Lyr stars of the OGLE Galactic bulge collection, of which 38 were originally classified as RRab and 4 as RRc, additional variability was detected. Its period is always shorter than period of the dominant variability. Period ratios are in the $(0.68,\,0.72)$ range. In the Petersen diagram in Fig.~\ref{fig:pet} these stars are marked with green open triangles. Additional variability is of relatively high amplitude, typically 20 per cent of the amplitude of the dominant variability. In 10 stars (24 per cent) Blazhko-type modulation was detected. The most interesting and distinct feature of the discussed group is the shape of the light curve of the dominant variability -- in all stars the light curve is a smooth triangular shape, void of any sharp features, as illustrated in Fig.~\ref{fig:lc} (bottom left). The corresponding Fourier parameters are in between those characteristic for single-periodic RRab stars and single-periodic RRc stars (Fig.~\ref{fig:lc}, right panels).

The nature of this group remains unknown. The characteristic triangular shape of the light curve suggests that dominant variability is due to radial fundamental mode. The additional variability cannot correspond to the radial first overtone then, at least assuming that physical parameters of these variables are similar to that of RR~Lyr stars or of RR~Lyr impostors -- Binary Evolution Pulsators, see e.g. \cite{bep_smolec}. 

\section{Concluding remarks}\label{sec:summary}

Over the recent years we witness the revival of classical pulsators. New dynamical effects, low-amplitude instabilities or new forms of multi-periodic pulsation were discovered. In the Petersen diagram the revolution took place -- ten years ago, the only group in the case of RR~Lyr stars were RRd variables. The radial mode paradigm for classical pulsators no longer holds -- non-radial modes are excited in at least one of the just described group, with additional periodicities in the $(0.60,0.65)\Po$ range. Their identification as $\ell\!=\!7$, $8$ and $9$ modes, might open the new window of asteroseismic investigations of classical pulsators. For a few other groups, the nature of additional periodicities remains unknown. Fortunately, the additional periodicities are either rare or of low amplitude. Hence, they have no practical impact on the use of classical pulsators for distance determination. It is disturbing however that we are far from complete understanding of the most useful astrophysical tools.

\begin{acknowledgement} 
\noindent\vskip 0.2cm
\noindent {\em Acknowledgments}: This research is supported by the Polish National Science Centre through grant DEC-2015/17/B/ST9/03421. HN is supported by the Polish Ministry of Science and Higher Education through grant no. 0192/DIA/2016/45. MS acknowledges financial support from the Hungarian NKFIH Grant K-115709.
\end{acknowledgement}

%
%

\end{document}